\documentclass[11pt,a4paper,psamsfonts,twoside,reqno]{amsart}
\usepackage[latin1]{inputenc}
\usepackage{amssymb,amsthm}
\usepackage[mathscr]{eucal}
\usepackage{mathrsfs}
\usepackage{graphicx,pstricks}
\usepackage{scalefnt}

\theoremstyle{plain}
\newtheorem{Theorem}{Theorem}[section]

\theoremstyle{definition}
\newtheorem{Definition}[Theorem]{Definition}
\newtheorem{Example}[Theorem]{Example}

\theoremstyle{remark}
\newtheorem{Remark}[Theorem]{Remark}

\begin{document}
\title{On the normal ordering of multi-mode boson operators}
\author{Toufik Mansour}
\address{Department of Mathematics, University of Haifa, Haifa 31905, Israel}
\email{toufik@math.haifa.ac.il}
\author{Matthias Schork}
\address{Alexanderstr. 76\\ 60489 Frankfurt, Germany}
\email{mschork@member.ams.org}
\date{\today}
%\subjclass{40A05,40A25}
\abstract In this article combinatorial aspects of normal ordering
annihilation and creation operators of a multi-mode boson system are
discussed. The modes are assumed to be coupled since otherwise the
problem of normal ordering is reduced to the corresponding problem
of the single-mode case. To describe the normal ordering in the
multi-mode case for each mode a colour is introduced and coloured
contractions are considered. A depiction for coloured contractions
via coloured linear representations is given. In analogy to the
single-mode case associated coloured Stirling numbers are defined as
coefficients appearing in the process of normal ordering powers of
the number operators. Several properties of these coloured Stirling
numbers are discussed.
\endabstract
\maketitle

\section{Introduction}
Since the seminal work of Katriel \cite{kat} the combinatorial
aspects of normal ordering arbitrary words in the creation and
annihilation operators $b^{\dag}$ and $b$ of a single-mode boson
having the usual commutation relations
\begin{equation}\label{commb}
[b,b^{\dag}]\equiv bb^{\dag}-b^{\dag}b=1,\hspace{0,3cm}[b,b]=0,\hspace{0,3cm}[b^{\dag},b^{\dag}]=0
\end{equation}
have been studied intensively and many generalizations (e.g., $q$-deformed bosons) have also been considered, see, e.g., \cite{wit0,mik,kk,kat2,kat3,blas,blas15,scho,blas2,fuji,blas3,blas35,blas4,varv,wit,scho2,ts,tsm} and the references given therein. For an important discussion and references to the earlier literature on normal ordering noncommuting operators see Wilcox \cite{wilc}. For the creation and annihilation operator $f^{\dag}$ and $f$ of a single-mode fermion having the usual anticommutation relations
\[
\{f,f^{\dag}\}\equiv ff^{\dag}+f^{\dag}f=1,\hspace{0,3cm}\{f,f\}=0,\hspace{0,3cm}\{f^{\dag},f^{\dag}\}=0
\]
the analogous combinatorial problem does not exist due to the nilpotency of the operators, i.e., $(f^{\dag})^2=0=f^2$. However, if one considers instead of a single-mode fermion a multi-mode fermion (i.e., several sets of operators $f_i,f_i^{\dag}$) then interesting combinatorial connections to rook numbers exist (note that the general normal ordering problem of a {\it single-mode} boson also has connections to rook numbers \cite{varv}). This was noted by Navon \cite{nav} even before Katriel demonstrated that normal ordering powers of the bosonic number operator, i.e., $(b^{\dag}b)^n$, involves the Stirling numbers of second kind \cite{kat}. To the best of our knowledge the normal ordering of creation and annihilation operators of a {\it multi-mode boson} has not been treated in the literature in a combinatorial fashion analogous to the single-mode case, although multi-mode systems are of particular importance in applications, e.g., in quantum optics \cite{barnett}. Clearly, to obtain interesting results the different modes have to be coupled (i.e., have a nontrivial commutation relation), otherwise the problem reduces to the simultanous normal ordering of the individual modes. In the present paper we assume that all modes $b_i^{\dag},b_i$ satisfy the commutation relations (\ref{commb}) and are coupled via $[b_i,b_j^{\dag}]=1$ (all other commutators between operators of different modes vanish). In this paper we begin the combinatorial study of the normal ordering of this multi-mode boson. It is important to note that the situation is slightly different from the one typically considered in quantum optics: There the different modes commute and the nontrivial interaction is introduced via the Hamiltonian; here we don't consider Hamiltonians but only the coupled creation and annihilation operators. Also, we discuss the relations coming from the algebra itself and consider only casually representations of the algebra.   

The paper is structured as follows. In Section 2 we discuss the two-mode boson, in particular the normal ordering problem. It is shown that the general normal ordering can be described with the help of the double dot operation and two-coloured contractions, a generalization of the coventional contractions. Some examples are also considered and the relation to the normal ordering process of a single-mode boson is discussed. In Section 3 we describe a graphical representation of two-coloured contractions using two-coloured linear representations. In Section 4 we introduce two-coloured Stirling numbers as coefficients appearing in the normal ordering process of the two-mode boson. Several properties of these two-coloured Stirling numbers are discussed, e.g., recursion relations and some explicit values. In Section 5 we discuss representations of the commutation relations of the two-mode boson and give a simple example. In Section 6 it is described how the results of the preceding sections can be generalized to the general multi-mode boson. Finally, in Section 7 some conclusions are presented.

\section{The two-mode boson}
In this section we discuss the two-mode boson. That means that we consider two sets $\{b^{\dag},b\}$ and $\{a^{\dag},a\}$ of bosonic creation and annihilation operators which each satisfy (\ref{commb}) and are in addition coupled via the commutation relations
\begin{equation}\label{commb2}
[a,b^{\dag}]=1,\hspace{0,3cm}[a,b]=0
\end{equation}
(together with the adjoint equations $[b,a^{\dag}]=1$ and $[a^{\dag},b^{\dag}]=0$). Since we will often compare this two-mode boson with the case of two noncoupled bosons it is convenient to introduce a terminology which will be used in the following.
\begin{Definition}
The {\it coupled two-mode boson} (or, briefly, {\it two-mode
boson}) consists of the operators $\{a,a^{\dag},b,b^{\dag}\}$
satisfying the commutation relations
\begin{equation}\label{inttwobos}
[a,a]=[a^{\dag},a^{\dag}]=[b,b]=[b^{\dag},b^{\dag}]=[a,b]=[a^{\dag},b^{\dag}]=0,
[a,a^{\dag}]=[b,b^{\dag}]=[a,b^{\dag}]=[b,a^{\dag}]=1.
\end{equation}
Similarly, the {\it noncoupled two-mode boson} consists of the operators $\{a,a^{\dag},b,b^{\dag}\}$ satisfying the commutation relations
\begin{equation}\label{ninttwobos}
[a,a]=[a^{\dag},a^{\dag}]=[b,b]=[b^{\dag},b^{\dag}]=[a,b]=[a^{\dag},b^{\dag}]=[a,b^{\dag}]=[b,a^{\dag}]=0,[a,a^{\dag}]=[b,b^{\dag}]=1.
\end{equation}
In both cases we speak of the modes $a$ and $b$ (consisting of the operators $\{a,a^{\dag}\}$ resp. $\{b,b^{\dag}\}$) and also imagine that the modes have different {\it colours}.
\end{Definition}

In the following we are interested in properties of the coupled two-mode boson and - if not stated otherwise - the commutation relations (\ref{inttwobos}) are assumed to hold.

Let us introduce the number operators $N_a$ and $N_b$ as well as the number-difference operator $N_d$ by
\begin{equation}
N_a :=a^{\dag}a, \hspace{0,3cm} N_b:=b^{\dag}b, \hspace{0,3cm} N_d:=N_a-N_b=a^{\dag}a-b^{\dag}b.
\end{equation}
It is straightforward to show that one has the usual relations $[N_a,a]=-a, [N_a,a^{\dag}]=a^{\dag}$ and $[N_b,b]=-b, [N_b,b^{\dag}]=b^{\dag}$ as well as the ``cross relations'' $[N_a,b]=-a, [N_a,b^{\dag}]=a^{\dag}$ and $[N_b,a]=-b, [N_a,b^{\dag}]=b^{\dag}$. Note that these relations imply that $a-b$ commutes with $N_a$ and with $N_b$, i.e., $[N_a,a-b]=0=[N_b,a-b]$. Also, one has $[N_a,N_b]=a^{\dag}b-b^{\dag}a$. The number-difference operator $N_d$ plays a prominent role in recent studies on superconductors, see, e.g., \cite{fan} (but note that the two modes discussed there commute in contrast to (\ref{commb2})).

The {\it normal ordering} is a functional representation of two-mode boson operator functions in which all the creation operators stand to the left of the annihilation operators. Although it is irrelevant from a strictly logical point of view (since $ab=ba$ as well as $a^{\dag}b^{\dag}=b^{\dag}a^{\dag}$) we will assume in the following that in the normally ordered form the operators $a^{\dag}$ precede the operators $b^{\dag}$ and that the operators $b$ precede the operators $a$. Let an arbitrary word $F(a,a^{\dag},b,b^{\dag})$ be given; a function $F(a,a^{\dag},b,b^{\dag})$ can be seen as a word on the alphabet $\{a,a^{\dag},b,b^{\dag}\}$. We denote
by $\mathcal{N}[F(a,a^{\dag},b,b^{\dag})]$ the normal ordering of the function $F(a,a^{\dag},b,b^{\dag})$.
Using the commutation relations (\ref{inttwobos}) it is clear that its normally ordered form $\mathcal{N}[F(a,a^{\dag},b,b^{\dag})]=F(a,a^{\dag},b,b^{\dag})$ can be written as
\begin{equation}
\mathcal{N}[F(a,a^{\dag},b,b^{\dag})]=F(a,a^{\dag},b,b^{\dag})=\sum_{k,l,m,n}C_{k,l,m,n}(a^{\dag})^k(b^{\dag})^lb^ma^n
\end{equation}
for some coefficients $C_{k,l,m,n}$ and the main task consists of determining the coefficients as explicit as possible.

\begin{Remark}\label{remark1} Let us consider the situation of the noncoupled case (\ref{ninttwobos}); this is just the two-dimensional harmonic oscillator and is the situation considered, e.g., in \cite{fan}. Given a function $F(a,a^{\dag},b,b^{\dag})$, we denote the function which results by deleting all operators $a,a^{\dag}$ by $F^{a-del}(b,b^{\dag})$; similarly, we denote the function which results by deleting all operators $b,b^{\dag}$ by $F^{b-del}(a,a^{\dag})$. Since all operators of the one mode commute with all operators of the other mode one has
\[
F(a,a^{\dag},b,b^{\dag})=F^{b-del}(a,a^{\dag})F^{a-del}(b,b^{\dag}).
\]
Clearly, the problem of normal ordering $F(a,a^{\dag},b,b^{\dag})$ reduces to the problem of normal ordering the single-mode operator functions $F^{b-del}(a,a^{\dag})$ and $ F^{a-del}(b,b^{\dag})$.
\end{Remark}

In the case of a single-mode boson one can describe the normal ordering by means of contractions and the double dot operation \cite{tsm}. Let us recall the procedure from \cite{tsm} for a word $F(b,b^{\dag})$. A \emph{contraction} consists
of substituting $b=\varnothing$ and $b^{\dagger}=\varnothing^{\dagger}$ in the
word whenever $b$ precedes $b^{\dagger}$ and deleting all the letters $\varnothing$ and
$\varnothing^{\dagger}$ in the word. Among all possible contractions we
also include the null contraction, that is the contraction leaving the word as
it is.
We define $\mathcal{C}(F(b,b^{\dagger}))$ to be the multiset of all contractions of $F(b,b^{\dagger})$.
The \emph{double dot operation} arranges a word $\pi \in \mathcal{C}(F(b,b^{\dagger}))$ such that all the
letters $b^{\dagger}$ precede the letters $b$ (e.g., $:b^{k}
(b^{\dagger})^{l}:$ $=(b^{\dagger})^{l}b^{k}$). Then one has
\begin{equation}\label{normal}
\mathcal{N}[F(b,b^{\dagger})]=F(b,b^{\dagger})=\sum_{\pi\in\mathcal{C}(F(b,b^{\dagger}))}:\pi:.
\end{equation}

We now introduce {\it two-coloured contractions} and show that in general the normal ordering of an operator function of the two-mode boson (\ref{inttwobos}) can be described with the help of two-coloured contractions and the {\it double dot operation}. Here the ``colours'' are used to mark the two different modes of operators involved and the use of ``colours'' will become clearer when we consider {\it two-coloured linear representations} in Section \ref{tclr}.

A \emph{two-coloured contraction} consists
of
\begin{enumerate}
\item substituting
\begin{enumerate}
\item $a=\varnothing$ and $a^{\dagger}=\varnothing^{\dagger}$ in the
word whenever $a$ precedes $a^{\dagger}$,
\item $b=\varnothing$ and $b^{\dagger}=\varnothing^{\dagger}$ in the
word whenever $b$ precedes $b^{\dagger}$,
\item $a=\varnothing$ and $b^{\dagger}=\varnothing^{\dagger}$ in the
word whenever $a$ precedes $b^{\dagger}$,
\item $b=\varnothing$ and $a^{\dagger}=\varnothing^{\dagger}$ in the
word whenever $b$ precedes $a^{\dagger}$,
\end{enumerate}
\item and deleting all the letters $\varnothing$ and
$\varnothing^{\dagger}$ in the word.
\end{enumerate}
Among all possible two-coloured contractions we
also include the null contraction, that is the contraction leaving the word as
it is. We define $\mathcal{C}^{(2)}(F(a,a^{\dag},b,b^{\dagger}))$ to be the multiset of all two-coloured contractions of $F(a,a^{\dag},b,b^{\dagger})$. Following \cite{tsm} we call a two-coloured contraction to be of {\it degree} $r$ if precisely $r$ pairs of creation and annihilation operators are contracted.

\begin{Example}
Let $F(a,a^{\dag},b,b^{\dag})=a^2a^{\dag}b^2b^{\dag}$. Since there are two creation operators in the word it is clear that contractions of degree at most two can arise. The null-contraction gives trivially $a^2a^{\dag}b^2b^{\dag}$. There do exist six contractions of degree one (here we have indicated in the brackets which two operators were contracted):
\[
ab^2b^{\dag}(13),aa^{\dag}b^2(16),ab^2b^{\dag}(23),aa^{\dag}b^2(26),a^2a^{\dag}b(46),a^2a^{\dag}b(56).
\]
Likewise, there do exist six contractions of degree two:
\[
b^2(13,26), ab(13,46), ab(13,56), b^2(16,23), ab(23,46), ab(23,56).
\]
Thus, the set of two-coloured contractions is given by
\begin{equation}\label{examplecol}
\begin{array}{l}
\mathcal{C}^{(2)}(a^2a^{\dag}b^2b^{\dag})\\
\qquad=\{a^2a^{\dag}b^2b^{\dag},ab^2b^{\dag},aa^{\dag}b^2,ab^2b^{\dag},aa^{\dag}b^2,a^2a^{\dag}b,a^2a^{\dag}b,b^2,b^2,
ab, ab, ab, ab\}. \end{array}
\end{equation}
\end{Example}

The {\it double dot operation} arranges a word $\pi \in \mathcal{C}^{(2)}(F(a,a^{\dag},b,b^{\dagger}))$ such that all the
letters $a^{\dag}$ precede all letters $b^{\dagger}$ precede all letters $b$ precede all letters $a$. As an example one has $:b^{k}a^m
(b^{\dagger})^{l}(a^{\dag})^n:$ $=(a^{\dag})^n(b^{\dagger})^{l}b^{k}a^m$. It is now possible to describe the generalization of (\ref{normal}) in the following theorem.

\begin{Theorem}\label{theorem1} Let $F(a,a^{\dag},b,b^{\dag})$ be an operator function of the two-mode boson (\ref{inttwobos}). The normally ordered form of $F(a,a^{\dag},b,b^{\dag})$ can be described with the help of two-coloured contractions and the double dot operation as follows:
\begin{equation}\label{normaltwo}
\mathcal{N}[F(a,a^{\dag},b,b^{\dag})]=F(a,a^{\dag},b,b^{\dag})=\sum_{\pi\in\mathcal{C}^{(2)}(F(a,a^{\dag},b,b^{\dagger}))}:\pi:.
\end{equation}
\end{Theorem}

\begin{Example}\label{exampleex}
Let $F(a,a^{\dag},b,b^{\dag})=a^2a^{\dag}b^2b^{\dag}$ from above. The associated set of two-coloured contractions is given by (\ref{examplecol}) and it follows from Theorem \ref{theorem1} that
\begin{equation}\label{exampleres}
\mathcal{N}[a^2a^{\dag}b^2b^{\dag}]= a^{\dag}b^{\dag}b^2a^2+2a^{\dag}ba^2+2a^{\dag}b^2a+2b^{\dag}b^2a+2b^2+4ba
\end{equation}
which can also be checked by hand using (\ref{inttwobos}). Let us now come back to Remark \ref{remark1} and assume that the two modes are not coupled, i.e., satisfy (\ref{ninttwobos}). It follows that $F^{a-del}(b,b^{\dag})=b^2b^{\dag}$ and $F^{b-del}(a,a^{\dag})=a^2a^{\dag}$ as well as $\mathcal{N}[a^2a^{\dag}b^2b^{\dag}]=\mathcal{N}[a^2a^{\dag}]\mathcal{N}[b^2b^{\dag}]=(a^{\dag}a^2+2a)(b^{\dag}b^2+2b)$. Multiplying out and sorting the terms yields
\begin{equation}\label{exampleresc}
\mathcal{N}[a^2a^{\dag}b^2b^{\dag}]=a^{\dag}b^{\dag}b^2a^2+2a^{\dag}ba^2+2b^{\dag}b^2a+4ba
\end{equation}
which clearly differs from the coupled case (\ref{exampleres}).
\end{Example}

\begin{Example}\label{excomm} In this example we again come back to Remark \ref{remark1} and assume that the two modes are not coupled, i.e., satisfy (\ref{ninttwobos}), implying that the modes comprise just the two-dimensional harmonic oscillator. This implies that $[N_a,N_b]=0$ which is crucial for the following. Let us consider the operator function $F(a,a^{\dag},b,b^{\dag})=N_a^{n_a}N_b^{n_b}$ for some powers $n_a,n_b\in \mathbf{N}$. It follows that $\mathcal{N}[N_a^{n_a}N_b^{n_b}]=\mathcal{N}[N_a^{n_a}]\mathcal{N}[N_b^{n_b}]$ and we may use the beautiful result of Katriel \cite{kat} that
\begin{equation}\label{katriel}
\mathcal{N}[N_a^{n_a}]=\sum_{k=0}^{n_a}S(n_a,k)(a^{\dag})^ka^k
\end{equation}
(where the $S(n_a,k)$ are the Stirling numbers of second kind) to conclude that
\begin{equation}\label{Stirlingtwo}
\mathcal{N}[N_a^{n_a}N_b^{n_b}]=\sum_{k=0}^{n_a}\sum_{l=0}^{n_b}S(n_a,k)S(n_b,l)(a^{\dag})^k(b^{\dag})^lb^la^k.
\end{equation}
Under these circumstances it is also easy to determine the normally ordered form of powers of the number-difference operator since $N_d=N_a-N_b$ and, therefore, $N_d^{n_d}=\sum_{k=0}^{n_d}{n_d \choose k}(-1)^kN_a^{n_d-k}N_b^k$, implying
\begin{equation}
\mathcal{N}[N_d^{n_d}]=\sum_{k=0}^{n_d}\sum_{r=0}^{n_d-k}\sum_{s=0}^{k}{n_d \choose k}(-1)^kS(n_d-k,r)S(k,s)(a^{\dag})^r(b^{\dag})^sb^sa^r.
\end{equation}
Given such a two-dimensional harmonic oscillator, a well-known application is to use it for a representation of the angular momentum operators. This is the {\it coupled boson representation} introduced by Schwinger \cite{schwing}, see also \cite{wit0,wilc} for discussions. Namely, if
\[
J_+:=b^{\dag}a, \hspace{0,3cm}J_-:=a^{\dag}b, \hspace{0,3cm}J_z:=\frac{1}{2}(N_b-N_a)=-\frac{1}{2}N_d
\]
then $J_+$ and $J_-$ are the angular momentum raising and lowering operators, respectively (and $J_z$ is the $z$-component).
\end{Example}

\begin{Example} Let us consider $F(a,a^{\dag},b,b^{\dag})=(b^ra^{\dag}a)^n=(b^rN_a)^n$. Its normally ordered form can be written as
\begin{equation}
\mathcal{N}[(b^ra^{\dag}a)^n]=\sum_{i,j,k}c_n^{(r)}(i,j,k)(a^{\dag})^ib^ja^k
\end{equation}
for some coefficients $c_n^{(r)}(i,j,k)$. Using $a^ka^{\dag}=a^{\dag}a^k+ka^{k-1}$, one shows
\[
(a^{\dag})^ib^ja^kb^ra^{\dag}a=(a^{\dag})^{i+}b^{j+r}a^{k+1}+(j+r)(a^{\dag})^ib^{j+r-1}a^{k+1}+k(a^{\dag})^ib^{j+r}a^k,
\]
implying the following recursion relation for the $c_n^{(r)}(i,j,k)$:
\begin{equation}
c_{n+1}^{(r)}(i,j,k)=c_n^{(r)}(i-1,j-r,k-1)+(j+r)c_n^{(r)}(i,j-r+1,k-1)+kc_n^{(r)}(i,j-r,k).
\end{equation}
Let us check that this recursion relation reduces for the special case $r=0$ to the one of the Stirling numbers of second kind (if $r=0$ then $F(a,a^{\dag},b,b^{\dag})=(a^{\dag}a)^n=N_a^n$ is the function of a single-mode boson considered in Example \ref{excomm}). For $r=0$ one has the recursion relation
\[
c_{n+1}^{(0)}(i,j,k)=c_n^{(0)}(i-1,j,k-1)+jc_n^{(0)}(i,j+1,k-1)+kc_n^{(0)}(i,j,k)
\]
where $c_{n}^{(0)}(i,j,k)=0$ for $j>0$. Thus, defining first $T_n(i,k):=c_{n}^{(0)}(i,0,k)$, this recursion relation reduces to $T_{n+1}(i,k)=T_n(i-1,k-1)+kT_n(i,k)$. From the above definition it is clear that $T_n(i,k)=0$ if $i\neq k$. Thus, only the case $i=k$ remains to be considered. Defining $\mathcal{S}(n,k):=T_n(k,k)=c_{n}^{(0)}(k,0,k)$, the last recursion relation reduces to
\begin{equation}\label{recstirl}
\mathcal{S}(n+1,k)=\mathcal{S}(n,k-1)+k\mathcal{S}(n,k)
\end{equation}
which is indeed the recursion relation for the Stirling numbers of second kind. The first few explicit values for the coefficients $c_{n}^{(r)}(i,j,k)$ are given by
\[
c_{0}^{(r)}(i,j,k)=\delta_{i,0}\delta_{j,0}\delta_{k,0},
\]
the only nonvanishing coefficients for $n=1$ are
\[
c_{1}^{(r)}(0,r-1,1)=r, \hspace{0,3cm}c_{1}^{(r)}(1,r,1)=1,
\]
and the only nonvanishing coefficients for $n=2$ are
\begin{eqnarray*}
c_{2}^{(r)}(2,2r,2)&=&1, \hspace{0,3cm} c_{2}^{(r)}(1,2r-1,2)=3r, \hspace{0,3cm}c_{2}^{(r)}(1,2r,1)=1, \\ c_{2}^{(r)}(0,2r-1,1)&=&r, \hspace{0,3cm}c_{2}^{(r)}(0,2r-2,2)=r(2r-1).
\end{eqnarray*}
\end{Example}

Given an operator function $F(a,a^{\dag},b,b^{\dag})$, one can associate to it an operator function of a single-mode boson by neglecting the colour, i.e., by ``putting the sunglasses on''. To be more precise, we consider on the one hand the unital $*$-algebra $\mathcal{A}^{(2)}$ generated by the elements $\{1,a,a^{\dag},b,b^{\dag}\}$ having the relations (\ref{inttwobos}) (and the involution $\dag$) as well as the algebra $\mathcal{A}^{(1)}$ generated by the elements $\{1,c,c^{\dag}\}$ having the relations (\ref{commb}). Of course, $\mathcal{A}^{(1)}$ is just the usual Heisenberg-Weyl-algebra. ``Putting the sunglasses on'' means introducing a $*$-algebra-homomorphism
\[
\mathcal{SUN}:\mathcal{A}^{(2)}\rightarrow \mathcal{A}^{(1)},
\]
mapping $1 \mapsto 1$ as well as $a^{\dag},b^{\dag}\mapsto c^{\dag}$ and $a,b \mapsto c$. Given an operator function $F(a,a^{\dag},b,b^{\dag}) \in \mathcal{A}^{(2)}$, we can either first put the sunglasses on and then normal order the resulting operator function of a single-mode boson, or we can first normal order it and then put the sunglasses on and obtain a normally ordered single-mode operator function. It is a beautiful fact that these two results agree, i.e., the order of the two operations is irrelevant.

\begin{Theorem}\label{sunglass2}
Let $F(a,a^{\dag},b,b^{\dag}) \in \mathcal{A}^{(2)}$ be given and let us denote the operation of normal ordering in the two algebras by $\mathcal{N}_i:\mathcal{A}^{(i)}\rightarrow \mathcal{A}^{(i)}$ for $i=1,2$. Then one has
\[
\mathcal{SUN}\{\mathcal{N}_2[F(a,a^{\dag},b,b^{\dag})]\}=\mathcal{N}_1[\mathcal{SUN}\{F(a,a^{\dag},b,b^{\dag})\}],
\]
or, in a more fancy language,
\begin{equation}\label{commmap}
\mathcal{SUN}\circ \mathcal{N}_2 =\mathcal{N}_1\circ\mathcal{SUN}.
\end{equation}
\end{Theorem}
However, if we want to reduce the combinatorial complexity of normal ordering the two-mode boson by mapping it to a single-mode boson and normal ordering there, we have to keep track of the original colours somehow. By this we mean that ``morally'' one wants to write $\mathcal{N}_2 =\mathcal{SUN}^{-1}\circ\mathcal{N}_1\circ\mathcal{SUN}$ (which, of course, makes no sense since $\mathcal{SUN}$ is not invertible).

\begin{Example} Let us continue our study of $F(a,a^{\dag},b,b^{\dag})=a^2a^{\dag}b^2b^{\dag} \in \mathcal{A}^{(2)}$ from above. ``Putting the sunglasses on'' yields $\mathcal{SUN}(a^2a^{\dag}b^2b^{\dag})=c^2c^{\dag}c^2c^{\dag}\in  \mathcal{A}^{(1)}$. A straightforward calculation shows that normal ordering this expression gives $\mathcal{N}_1[\mathcal{SUN}(a^2a^{\dag}b^2b^{\dag})]=(c^{\dag})^2c^4+6c^{\dag}c^3+6c^2$. On the other hand, we have calculated the normally ordered form of $a^2a^{\dag}b^2b^{\dag}$, i.e., $\mathcal{N}_2[a^2a^{\dag}b^2b^{\dag}] \in \mathcal{A}^{(2)}$, in Example \ref{exampleex} with the result (\ref{exampleres}). Applying $\mathcal{SUN}$ to the right-hand side of (\ref{exampleres}) gives $\mathcal{SUN}(\mathcal{N}_2[a^2a^{\dag}b^2b^{\dag}])=(c^{\dag})^2c^4+6c^{\dag}c^3+6c^2$ which agrees with $\mathcal{N}_1[\mathcal{SUN}(a^2a^{\dag}b^2b^{\dag})]$ - as it should according to Theorem \ref{sunglass2}. Note that the above-mentioned problem with writing $\mathcal{N}_2 =\mathcal{SUN}^{-1}\circ\mathcal{N}_1\circ\mathcal{SUN}$, i.e., with the ``map'' $\mathcal{SUN}^{-1}$, amounts to the problem of reconstructing the right-hand side of (\ref{exampleres}) from $(c^{\dag})^2c^4+6c^{\dag}c^3+6c^2$ - which is not possible without further information.
\end{Example}

\begin{Remark}
It is important that in $\mathcal{A}^{(2)}$ the modes are coupled, i.e., that $[a,b^{\dag}]=1=[b,a^{\dag}]$ holds true since otherwise Theorem \ref{sunglass2} - in particular (\ref{commmap}) - would not be true. Note that on the right-hand side of (\ref{commmap}) the relation between the two modes $a$ and $b$ is irrelevant whereas it is crucial for the left-hand side. Thus, for the equation to hold one cannot choose an arbitrary relation between the two modes. As an example, we consider again the function $F(a,a^{\dag},b,b^{\dag})=a^2a^{\dag}b^2b^{\dag}$ and assume that the two modes are not coupled (all operators of one mode commute with all operators of the other mode, i.e., (\ref{ninttwobos}) holds true). Using (\ref{exampleresc}), we find that $\mathcal{SUN}(\mathcal{N}_2[a^2a^{\dag}b^2b^{\dag}])=(c^{\dag})^2c^4+4c^{\dag}c^3+4c^2\neq \mathcal{N}_1[\mathcal{SUN}(a^2a^{\dag}b^2b^{\dag})]$.
\end{Remark}

\section{Two-coloured linear representations}\label{tclr}
Let us recall from \cite{tsm} that contractions in the single-mode case can be depicted with diagrams called {\it linear representations}. Let us consider a word $\pi$ on the alphabet $\{b,b^{\dag}\}$ of length $m$, i.e., $\pi=\pi_1\cdots \pi_m$ with $\pi_j\in \{b,b^{\dag}\}$. We
draw $m$ vertices, say $1,2,\ldots,m$, on a horizontal line, such that the
point $j$ corresponds to the letter $\pi_{j}$. We represent each $b$ by a
white vertex and each letter $b^{\dagger}$ by a black vertex; a black
vertex $j$ can be connected by an undirected edge $(i,j)$ to a white vertex $i$ if $i<j$ (but there may also be black vertices having no edge).
Importantly, the edges are drawn in the plane above the points. This is the linear representation of a contraction.

Now, we generalize this to the coupled two-mode case (\ref{inttwobos}) by introducing {\it two-coloured linear representations}. We assume that two colours $c_a$ and $c_b$ have been chosen which are associated to the modes $a$ and $b$. Let us consider a word $\pi$ on the alphabet $\{a,a^{\dag},b,b^{\dag}\}$ of length $m$, i.e., $\pi=\pi_1\cdots \pi_m$ with $\pi_j\in \{a,a^{\dag},b,b^{\dag}\}$. We
draw $m$ vertices, say $1,2,\ldots,m$, on a horizontal line, such that the
point $j$ corresponds to the letter $\pi_{j}$. We represent
\begin{enumerate}
\item each letter $a$ by an empty circle of colour $c_a$,
\item each letter $a^{\dagger}$ by a filled circle of colour $c_a$,
\item each letter $b$ by an empty circle of colour $c_b$,
\item each letter $b^{\dagger}$ by a filled circle of colour $c_b$.
\end{enumerate}
A filled circle $j$ can be connected by an undirected edge $(i,j)$ to an empty circle $i$ if $i<j$ - independent of the colours of the circles. As above, the edges are drawn in the plane above the points. This is the two-coloured linear representation of a two-coloured contraction.

\begin{Example}
The two-coloured linear representations for the function $F(a,a^{\dag},b,b^{\dag})=a^2a^{\dag}b^2b^{\dag}$ from above are shown in Figure \ref{falt}. Note that we have - due to technical reasons - represented the different colours by different symbols: The mode $a$ is represented by empty of filled black boxes, whereas the mode $b$ is represented by empty or filled circles.

\begin{figure}[h]
\begin{center}
\begin{pspicture}(0,0)(10,.4)
\setlength{\unitlength}{3mm} \linethickness{0.3pt}
\multips(-1,0)(3.4,0){4}{\psframe(0,-.05)(0.1,.05)\psframe(0.5,-.05)(0.6,.05)\psframe*(1,-.05)(1.1,.05)\pscircle(1.5,0){.2}\pscircle(2.0,0){.2}\pscircle*(2.5,0){.2}}
\linethickness{0.8pt}
\qbezier(8.1,.1)(9.7,1)(11.3,.1)
\qbezier(19.4,.1)(23.55,1)(27.7,.1)
\qbezier(32.4,.1)(33.15,1)(33.9,.1)
\put(-3.6,-1.2){$1$}\put(-1.9,-1.2){$2$}\put(-.2,-1.2){$3$}\put(1.5,-1.2){$4$}\put(3.2,-1.2){$5$}\put(4.9,-1.2){$6$}
\put(7.7,-1.2){$1$}\put(9.4,-1.2){$2$}\put(11.1,-1.2){$3$}\put(12.8,-1.2){$4$}\put(14.5,-1.2){$5$}\put(16.2,-1.2){$6$}
\put(19,-1.2){$1$}\put(20.7,-1.2){$2$}\put(22.4,-1.2){$3$}\put(24.1,-1.2){$4$}\put(25.8,-1.2){$5$}\put(27.5,-1.2){$6$}
\put(30.3,-1.2){$1$}\put(32,-1.2){$2$}\put(33.7,-1.2){$3$}\put(35.4,-1.2){$4$}\put(37.1,-1.2){$5$}\put(38.8,-1.2){$6$}
\end{pspicture}
\par
\begin{pspicture}(0,0)(10,1)
\setlength{\unitlength}{3mm} \linethickness{0.3pt}
\multips(-1,0)(3.4,0){4}{\psframe(0,-.05)(0.1,.05)\psframe(0.5,-.05)(0.6,.05)\psframe*(1,-.05)(1.1,.05)\pscircle(1.5,0){.2}\pscircle(2.0,0){.2}\pscircle*(2.5,0){.2}}
\linethickness{0.8pt}
\qbezier(-1.5,.1)(1.8,1)(5.1,.1)
\qbezier(13,.1)(14.7,1)(16.4,.1)
\qbezier(26,.1)(26.85,1)(27.7,.1)
\qbezier(30.7,.1)(32.3,1)(33.9,.1)\qbezier(32.4,.1)(35.7,1)(39,.1)
\put(-3.6,-1.2){$1$}\put(-1.9,-1.2){$2$}\put(-.2,-1.2){$3$}\put(1.5,-1.2){$4$}\put(3.2,-1.2){$5$}\put(4.9,-1.2){$6$}
\put(7.7,-1.2){$1$}\put(9.4,-1.2){$2$}\put(11.1,-1.2){$3$}\put(12.8,-1.2){$4$}\put(14.5,-1.2){$5$}\put(16.2,-1.2){$6$}
\put(19,-1.2){$1$}\put(20.7,-1.2){$2$}\put(22.4,-1.2){$3$}\put(24.1,-1.2){$4$}\put(25.8,-1.2){$5$}\put(27.5,-1.2){$6$}
\put(30.3,-1.2){$1$}\put(32,-1.2){$2$}\put(33.7,-1.2){$3$}\put(35.4,-1.2){$4$}\put(37.1,-1.2){$5$}\put(38.8,-1.2){$6$}
\end{pspicture}
\par
\begin{pspicture}(0,0)(10,1)
\setlength{\unitlength}{3mm} \linethickness{0.3pt}
\multips(-1,0)(3.4,0){4}{\psframe(0,-.05)(0.1,.05)\psframe(0.5,-.05)(0.6,.05)\psframe*(1,-.05)(1.1,.05)\pscircle(1.5,0){.2}\pscircle(2.0,0){.2}\pscircle*(2.5,0){.2}}
\linethickness{0.8pt}
\qbezier(-3.2,.1)(-1.8,1)(0,.1)\qbezier(1.7,.1)(3.4,1)(5.1,.1)
\qbezier(8.1,.1)(9.7,1)(11.3,.1)\qbezier(14.7,.1)(15.55,1)(16.4,.1)
\qbezier(19.4,.1)(23.55,1.2)(27.7,.1)\qbezier(21.1,.1)(21.85,0.8)(22.6,.1)
\qbezier(32.4,.1)(33.15,1)(33.9,.1)\qbezier(35.6,.1)(37.3,1)(39,.1)
\put(-3.6,-1.2){$1$}\put(-1.9,-1.2){$2$}\put(-.2,-1.2){$3$}\put(1.5,-1.2){$4$}\put(3.2,-1.2){$5$}\put(4.9,-1.2){$6$}
\put(7.7,-1.2){$1$}\put(9.4,-1.2){$2$}\put(11.1,-1.2){$3$}\put(12.8,-1.2){$4$}\put(14.5,-1.2){$5$}\put(16.2,-1.2){$6$}
\put(19,-1.2){$1$}\put(20.7,-1.2){$2$}\put(22.4,-1.2){$3$}\put(24.1,-1.2){$4$}\put(25.8,-1.2){$5$}\put(27.5,-1.2){$6$}
\put(30.3,-1.2){$1$}\put(32,-1.2){$2$}\put(33.7,-1.2){$3$}\put(35.4,-1.2){$4$}\put(37.1,-1.2){$5$}\put(38.8,-1.2){$6$}
\end{pspicture}
\par
\begin{pspicture}(0,0)(10,1)
\setlength{\unitlength}{3mm} \linethickness{0.3pt}
\multips(-1,0)(0.1,0){1}{\psframe(0,-.05)(0.1,.05)\psframe(0.5,-.05)(0.6,.05)\psframe*(1,-.05)(1.1,.05)\pscircle(1.5,0){.2}\pscircle(2.0,0){.2}\pscircle*(2.5,0){.2}}
\linethickness{0.8pt}
\qbezier(-1.5,.1)(-0.75,1)(0,.1)\qbezier(3.4,.1)(4.25,1)(5.1,.1)
\put(-3.6,-1.2){$1$}\put(-1.9,-1.2){$2$}\put(-.2,-1.2){$3$}\put(1.5,-1.2){$4$}\put(3.2,-1.2){$5$}\put(4.9,-1.2){$6$}
\end{pspicture}
\end{center}
\caption{The two-coloured linear representations of the two-coloured contractions of the word
$aaa^{\dagger}bbb^{\dagger}$.}
\label{falt}
\end{figure}
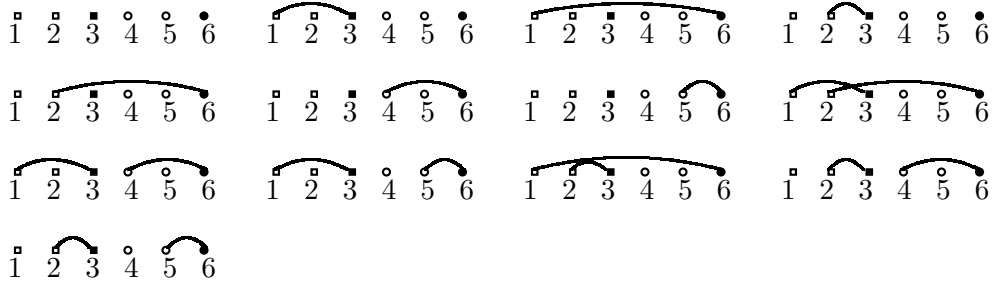
\end{Example}

\section{Two-coloured Stirling numbers}\label{sectwostirl}
In this section we introduce {\it two-coloured Stirling numbers} by generalizing the observation (\ref{katriel}) of Katriel \cite{kat}. The most natural analogue to (\ref{katriel}) is to consider the normally ordered form of either $N_a^{n_a}N_b^{n_b}$ or $(N_aN_b)^n$.
\begin{Definition} Let $n,n_a,n_b$ be natural numbers and let the operators $\{a,a^{\dag},b,b^{\dag}\}$ satisfy the commutation relations (\ref{inttwobos}) of the coupled two-mode boson. Then the {\it two-coloured Stirling numbers of first type} $S_{n_a,n_b}(i,j,k,l)$ are defined by
\begin{equation}
\mathcal{N}[N_a^{n_a}N_b^{n_b}]=\sum_{i,j,k,l}S_{n_a,n_b}(i,j,k,l)(a^{\dag})^i(b^{\dag})^jb^ka^l
\end{equation}
and the {\it two-coloured Stirling numbers of second type} $S_n(i,j,k,l)$ are defined by
\begin{equation}
\mathcal{N}[(N_aN_b)^n]=\sum_{i,j,k,l}S_n(i,j,k,l)(a^{\dag})^i(b^{\dag})^jb^ka^l.
\end{equation}
\end{Definition}

\begin{Remark}
Let us come back to Example \ref{excomm}. Assuming that the modes are not coupled, i.e. (\ref{ninttwobos}) holds true, we have determined $\mathcal{N}[N_a^{n_a}N_b^{n_b}]$ in (\ref{Stirlingtwo}), implying that in the noncoupled case
\[
S_{n_a,n_b}(i,j,k,l)=S(n_a,i)S(n_b,j)\delta_{i,l}\delta_{j,k}.
\]
\end{Remark}

Let us start with the two-coloured Stirling numbers of first type. It is clear that these numbers reduce to the conventional Stirling numbers of second kind if $n_a=0$ or $n_b=0$. More precisely, if $n_b=0$ then $j=0=k$ and $i=l$ and we have the identification $S_{n_a,0}(i,0,0,i)\equiv S(n_a,i)$. Similarly, if $n_a=0$ then $i=0=l$ and $j=k$ and we have the identification $S_{0,n_b}(0,j,j,0)\equiv S(n_b,j)$. From the definition it is clear that due to the two-coloured contractions the number of creation and annihilation numbers are equal for all summands, i.e., only summands with $i+j=k+l$ are not vanishing. An expansion shows that $N_a^{n_a}N_b^{n_b}=(a^{\dag})^{n_a}(b^{\dag})^{n_b}b^{n_b}a^{n_a}+\cdots$ where the dots mean lower order terms. Thus, $S_{n_a,n_b}(n_a,n_b,n_b,n_a)=1$ and $S_{n_a,n_b}(i,j,k,l)=0$ if $i,l>n_a$ or $j,k>n_b$. Let us consider some explicit examples for small exponents, beginning with $n_a=1=n_b$. One has
\[
N_aN_b=a^{\dag}b^{\dag}ba+a^{\dag}b,
\]
so that the only nonvanishing Stirling numbers $S_{1,1}(i,j,k,l)$ are
\[
S_{1,1}(1,1,1,1)=1, \hspace{0,3cm}S_{1,1}(1,0,1,0)=1.
\]
Similarly, one has
\[
N_a^2N_b=(a^{\dag})^2b^{\dag}ba^2+2(a^{\dag})^2ba+a^{\dag}b^{\dag}ba+a^{\dag}b,
\]
so that the only nonvanishing Stirling numbers $S_{2,1}(i,j,k,l)$ are
\[
S_{2,1}(2,1,1,2)=1, \hspace{0,3cm}S_{2,1}(2,0,1,1)=2, \hspace{0,3cm}S_{2,1}(1,1,1,1)=1, \hspace{0,3cm}
S_{2,1}(1,0,1,0)=1.
\]
Since
\[
N_aN_b^2=a^{\dag}(b^{\dag})^2b^2a+2a^{\dag}b^{\dag}b^2+a^{\dag}b^{\dag}ba+a^{\dag}b
\]
the only nonvanishing Stirling numbers $S_{1,2}(i,j,k,l)$ are
\[
S_{1,2}(1,2,2,1)=1, \hspace{0,3cm}S_{1,2}(1,1,2,0)=2, \hspace{0,3cm}S_{1,2}(1,1,1,1)=1, \hspace{0,3cm}
S_{1,2}(1,0,1,0)=1.
\]
Our final explicit example is given by
\begin{eqnarray*}
N_a^2N_b^2&=&(a^{\dag})^2(b^{\dag})^2b^2a^2+4(a^{\dag})^2b^{\dag}b^2a+(a^{\dag})^2b^{\dag}ba^2+2(a^{\dag})^2b^2\\&&+2(a^{\dag})^2ba+a^{\dag}(b^{\dag})^2b^2a +a^{\dag}b^{\dag}b^2+2a^{\dag}b^{\dag}ba+a^{\dag}b,
\end{eqnarray*}
showing that the only nonvanishing Stirling numbers
$S_{2,2}(i,j,k,l)$ are
$$\begin{array}{llll} S_{2,2}(2,2,2,2)=1,
& S_{2,2}(2,1,2,1)=4, & S_{2,2}(2,1,1,2)=1, &
S_{2,2}(2,0,2,0)=2, \\
S_{2,2}(2,0,1,1)=2, & S_{2,2}(1,2,2,1)=1, &S_{2,2}(1,1,2,0)=1, &
S_{2,2}(1,1,1,1)=2, \\
S_{2,2}(1,0,1,0)=1.
\end{array}$$
Using
\[
a^{\dag}a(a^{\dag})^i(b^{\dag})^jb^ka^l=(a^{\dag})^{i+1}(b^{\dag})^jb^ka^{l+1}+j(a^{\dag})^{i+1}(b^{\dag})^{j-1}b^ka^l+i(a^{\dag})^i(b^{\dag})^jb^ka^l,
\]
one finds the following recursion relation
\[
S_{n_a+1,n_b}(i,j,k,l)=S_{n_a,n_b}(i-1,j,k,l-1)+(j+1)S_{n_a,n_b}(i-1,j+1,k,l)+iS_{n_a,n_b}(i,j,k,l).
\]
Similarly, using
\[
(a^{\dag})^i(b^{\dag})^jb^ka^lb^{\dag}b=(a^{\dag})^i(b^{\dag})^{j+1}b^{k+1}a^l+k(a^{\dag})^i(b^{\dag})^jb^ka^l+l(a^{\dag})^i(b^{\dag})^jb^{k+1}a^{l-1},
\]
one finds the analogous recursion relation
\[
S_{n_a,n_b+1}(i,j,k,l)=S_{n_a,n_b}(i,j-1,k-1,l)+kS_{n_a,n_b}(i,j,k,l)+(l+1)S_{n_a,n_b}(i,j,k-1,l+1).
\]
Considering the first recursion relation for $n_b=0$ (with $j=0=k, i=l$ and identifying $S_{n_a,0}(i,0,0,i)\equiv S(n_a,i)$, see above), this reduces to $S(n_a+1,i)=S(n_a,i-1)+iS(n_a,i)$ which is the recursion relation (\ref{recstirl}) of the Stirling numbers of second kind. In the same fashion the second recursion relation reduces for $n_a=0$ also to the recursion relation of the Stirling numbers of second kind.

Let us now consider the two-coloured Stirling numbers of second type, $S_n(i,j,k,l)$. From the definition it is clear that $S_0(0,0,0,0)=1$ and that all other numbers with $n=0$ vanish. Considering $n=1$ shows that $S_1(i,j,k,l)=S_{1,1}(i,j,k,l)$. Thus, the first interesting case is $n=2$. Here one has
\begin{eqnarray*}
(N_aN_b)^2&=&(a^{\dag})^2(b^{\dag})^2b^2a^2+3(a^{\dag})^2b^{\dag}b^2a+(a^{\dag})^2b^{\dag}ba^2+(a^{\dag})^2b^2+(a^{\dag})^2ba\\&&+a^{\dag}(b^{\dag})^2b^2a +a^{\dag}(b^{\dag})^2ba^2+a^{\dag}b^{\dag}b^2+4a^{\dag}b^{\dag}ba+a^{\dag}b.
\end{eqnarray*}
Note that $(N_aN_b)^2\neq N_a^2N_b^2$. Thus, the only nonvanishing
Stirling numbers $S_{2}(i,j,k,l)$ are
$$\begin{array}{llll}
S_{2}(2,2,2,2)=1, & S_{2}(2,1,2,1)=3, & S_{2}(2,1,1,2)=1, &
S_{2}(2,0,2,0)=1, \\
S_{2}(2,0,1,1)=1, & S_{2}(1,2,2,1)=1, & S_{2}(1,2,1,2)=1, &
S_{2}(1,1,2,0)=1, \\
S_{2}(1,1,1,1)=4, & S_{2}(1,0,1,0)=1.
\end{array}$$
From their definition it is also clear that $S_{n}(i,j,k,l)$ vanishes if $i+j\neq k+l$. We now summarize the properties of the two-coloured Stirling numbers in the following theorem.
\begin{Theorem}\label{rec} The two-coloured Stirling numbers of first type $S_{n_a,n_b}(i,j,k,l)$ satisfy $S_{0,0}(i,j,k,l)=\delta_{i,0}\delta_{j,0}\delta_{k,0}\delta_{l,0}$ and the following recursion relation with respect to $n_a$
\[
S_{n_a+1,n_b}(i,j,k,l)=S_{n_a,n_b}(i-1,j,k,l-1)+(j+1)S_{n_a,n_b}(i-1,j+1,k,l)+iS_{n_a,n_b}(i,j,k,l)
\]
and $n_b$
\[
S_{n_a,n_b+1}(i,j,k,l)=S_{n_a,n_b}(i,j-1,k-1,l)+kS_{n_a,n_b}(i,j,k,l)+(l+1)S_{n_a,n_b}(i,j,k-1,l+1).
\]
Furthermore, one has $S_{n_a,n_b}(n_a,n_b,n_b,n_a)=1$ and $S_{n_a,n_b}(i,j,k,l)=0$ unless $i,l\leq n_a, j,k\leq n_b$ and $i+j=k+l$. The two-coloured Stirling numbers of second type $S_{n}(i,j,k,l)$ satisfy $S_{0}(i,j,k,l)=\delta_{i,0}\delta_{j,0}\delta_{k,0}\delta_{l,0}$ and the recursion relation
\begin{eqnarray*}
S_{n+1}(i,j,k,l)&=&S_{n}(i-1,j-1,k-1,l-1)+(j+1)S_{n}(i-1,j,k-1,l)\\&&+jS_{n}(i-1,j,k,l-1)+(j+1)^2S_{n}(i-1,j+1,k,l)
+iS_{n}(i,j-1,k-1,l)\\&&+iS_{n}(i,j-1,k,l-1)+i(2j+1)S_{n}(i,j,k,l)+i(i+1)S_{n}(i+1,j-1,k,l).
\end{eqnarray*}
Furthermore, one has $S_{n}(n,n,n,n)=1$ and $S_{n}(i,j,k,l)=0$ unless $i,j,k,l\leq n$ and $i+j=k+l$.
\end{Theorem}
\begin{proof} The properties of the $S_{n_a,n_b}(i,j,k,l)$ were shown above. The only nontrivial property which remains to be shown is the recursion relation of the $S_{n}(i,j,k,l)$. A straightforward but tedious computation shows that
\begin{eqnarray*}
a^{\dag}ab^{\dag}b(a^{\dag})^i(b^{\dag})^jb^ka^l&=&(a^{\dag})^{i+1}(b^{\dag})^{j+1}b^{k+1}a^{l+1}+(j+1)(a^{\dag})^{i+1}(b^{\dag})^jb^{k+1}a^l+j(a^{\dag})^{i+1}(b^{\dag})^jb^ka^{l+1}\\ &&+ j^2(a^{\dag})^{i+1}(b^{\dag})^{j-1}b^ka^l +i(a^{\dag})^i(b^{\dag})^{j+1}b^{k+1}a^l+i(a^{\dag})^i(b^{\dag})^{j+1}b^ka^{l+1} \\ &&+i(2j+1)(a^{\dag})^i(b^{\dag})^jb^ka^l+i(i-1)(a^{\dag})^{i-1}(b^{\dag})^{j+1}b^ka^l.
\end{eqnarray*}
This yields directly the asserted recusion relation for the $S_{n}(i,j,k,l)$.
\end{proof}

The recurrence relations in the statement of Theorem~\ref{rec} can
be written by partial differential equations as follows. Let us define (as
we said $l=i+j-k$)
\begin{eqnarray*}
T_{n_a,n_b}(u,v,w)&:=&\sum_{i,j,k}S_{n_a,n_b}(i,j,k,i+j-k)u^iv^jw^k,\\
T_{n}(u,v,w)&:=&\sum_{i,j,k}S_{n}(i,j,k,i+j-k)u^iv^jw^k.
\end{eqnarray*}
Multiplying the recurrence relations in the statement of Theorem~\ref{rec} by
$u^iv^jw^k$ and summing over all possible $i,j,k\geq0$ we get that
$$\begin{array}{l}
T_{n_a+1,n_b}(u,v,w)=u\left(1+\frac{\partial}{\partial
u}+\frac{\partial}{\partial v}\right)T_{n_a,n_b}(u,v,w),\\
T_{n_a,n_b+1}(u,v,w)=w\left(v+u\frac{\partial}{\partial
u}+v\frac{\partial}{\partial v}+(1-w)\frac{\partial}{\partial w}\right)T_{n_a,n_b}(u,v,w),\\
T_{n+1}(u,v,w)=u\left[w(v+1)+(vw+v+1)\left(\frac{\partial}{\partial
u}+\frac{\partial}{\partial
v}\right)+v\left(\frac{\partial}{\partial
u}+\frac{\partial}{\partial v}\right)^2\right]T_{n}(u,v,w),
\end{array}$$
for all $n_a,n_b\geq1$ and $n\geq0$ with the initial conditions
$T_{1,1}(u,v,w)=uw(v+1)$ and $T_0(u,v,w)=1$. The above recurrence relation for $T_{n+1}(u,v,w)$ can be written as
follows. Let
\[T\equiv T(x;u,v,w):=\sum_{n\geq0}T_n(u,v,w)x^n.
\]
Multiplying the recurrence relation for $T_n(u,v,w)$ by $x^n$ and summing over all $n\geq0$ yields the rather complicated looking partial differential equation
$$T=1+xuw(v+1)T+xu(vw+v+1)\left(\frac{\partial T}{\partial u}+\frac{\partial T}{\partial v}\right)
+xuv\left(\frac{\partial^2 T}{\partial
u^2}+2\frac{\partial^2 T}{\partial u\partial
v}+\frac{\partial^2T}{\partial v^2}\right).$$

Recall that the conventional Bell numbers may be defined as a sum over the Stirling numbers of second kind, i.e., $B(n)=\sum_{k}S(n,k)$, and are given explicitly by $B(0)=1, B(1)=1, B(2)=2,B(3)=5, B(4)=15,\ldots$ After having introduced and discussed some properties of the two-coloured Stirling numbers we may also introduce the associated {\it two-coloured Bell numbers of first resp. second type} by
\[
B^{(2)}(n_a,n_b):=\sum_{i,j,k,l}S_{n_a,n_b}(i,j,k,l), \hspace{0,3cm}B^{(2)}(n):=\sum_{i,j,k,l}S_{n}(i,j,k,l).
\]
From the explicit values for the two-coloured Stirling numbers given above we find the explicit values
\begin{eqnarray*}
B^{(2)}(0,0)&=&1, \hspace{0,3cm}B^{(2)}(1,0)=1, \hspace{0,3cm}B^{(2)}(0,1)=1, \hspace{0,3cm}B^{(2)}(1,1)=2, \\ B^{(2)}(2,1)&=&5, \hspace{0,3cm}B^{(2)}(1,2)=5, \hspace{0,3cm}B^{(2)}(2,2)=15.
\end{eqnarray*}
Similarly, one has
\[
B^{(2)}(0)=1, \hspace{0,3cm}B^{(2)}(1)=2, \hspace{0,3cm}B^{(2)}(2)=15.
\]
Thus, a comparison with the above-mentioned values of the conventional Bell numbers suggests that $B^{(2)}(n_a,n_b)=B(n_a+n_b)$ as well $B^{(2)}(n)=B(2n)$. This is indeed true:

\begin{Theorem}\label{Bellcomp} The two-coloured Bell numbers $B^{(2)}(n_a,n_b)$ of first type and $B^{(2)}(n)$ of second type are given by the conventional Bell numbers $B(n)$ as follows:
\begin{equation}
B^{(2)}(n_a,n_b)=B(n_a+n_b),\hspace{0,3cm}B^{(2)}(n)=B(2n).
\end{equation}
\end{Theorem}
\begin{proof} Let us show this for the two-coloured Bell numbers of first type; the proof is the same for those of the second type. From the definition of the two-coloured Stirling numbers of first type one has $\mathcal{N}_2[N_a^{n_a}N_b^{n_b}]=\sum_{i,j,k,l}S_{n_a,n_b}(i,j,k,l)(a^{\dag})^i(b^{\dag})^jb^ka^l$ where we have indicated with the subsript the algebra in which we normal order. We now want to make use of Theorem \ref{sunglass2}. Let us first apply $\mathcal{SUN}$ to the above equation, resulting in
\[
\mathcal{SUN}\{\mathcal{N}_2[N_a^{n_a}N_b^{n_b}]\}=\sum_{i,j,k}S_{n_a,n_b}(i,j,k,i+j-k)(c^{\dag})^{i+j}c^{i+j}
\]
where we have denoted by $c,c^{\dag}$ the annihilation and creation operator of the single-mode case and have also used that $k+l=i+j$. Now, we use (\ref{commmap}) and write the left-hand side of this equation as $\mathcal{N}_1[\mathcal{SUN}\{N_a^{n_a}N_b^{n_b}\}]=\mathcal{N}_1[N_c^{n_a+n_b}]$. Using (\ref{katriel}), we have obtained
\[
\sum_{l}S(n_a+n_b,l)(c^{\dag})^{l}c^{l}
=\sum_{i,j,k}S_{n_a,n_b}(i,j,k,i+j-k)(c^{\dag})^{i+j}c^{i+j}.
\]
Setting $c=1=c^{\dag}$ yields the desired equation $B(n_a+n_b)=B^{(2)}(n_a,n_b)$.
\end{proof}

\begin{Remark} In this particular case the operation of ``putting the sunglasses on'' (i.e., mapping to the single-mode case via $\mathcal{SUN}$) proved to be successful - in contrast to the case of the two-coloured Stirling numbers. Clearly, the reason for this is that the subtle distribution of summands in the two-mode case - which gets lost upon maping to the single-mode case - does not matter here since a sum over all possible terms is taken.

\end{Remark}

\section{A representation of the commutation relations}
Let us recall that in the single-mode case one has the following representation of the canonical commutation relations (\ref{commb}): $a^{\dag}\mapsto x,a\mapsto \frac{d}{dx} $ (acting as multiplication resp. differentiation operator on an appropriate set of functions). Introducing for the second mode the corresponding representation $b^{\dag}\mapsto y,b\mapsto \frac{d}{dy} $ one obtains a representation of the noncoupled two-mode boson (\ref{ninttwobos}).

We now construct a representation of the coupled two-mode boson (\ref{inttwobos}). For this we make an ansatz
\begin{equation}
a=q(x,y)\frac{\partial}{\partial x}, \hspace{0,3cm}a^{\dag}=s(x,y),\hspace{0,3cm}b=p(x,y)\frac{\partial}{\partial y}, \hspace{0,3cm}b^{\dag}=t(x,y)
\end{equation}
where $p,q,s,t$ are in the moment arbitrary (smooth) functions and the operators act on a space $D$ of (smooth) functions $f(x,y)$ which is also to be determined. Inserting this ansatz into the commutation relations (\ref{inttwobos}) yields conditions on the functions of the ansatz. For example, the relation $[a,b]=0$ yields the condition
\[
0=[a,b]f=abf-baf=q\frac{\partial}{\partial x}\left(p\frac{\partial f}{\partial y}\right)-p\frac{\partial}{\partial y}\left(q\frac{\partial f}{\partial x}\right),
\]
and, therefore,
\[
p\frac{\partial q}{\partial y}\frac{\partial f}{\partial x}=q\frac{\partial p}{\partial x}\frac{\partial f}{\partial y}
\]
(where we have used $\frac{\partial^2f}{\partial x\partial y}=\frac{\partial^2f}{\partial y\partial x}$). In the same fashion one considers the remaining commutation relations of (\ref{inttwobos}). As a result, one has the following set of five equations for the functions $p,q,s,t$ and $f$:
\begin{equation}\label{cond}
q\frac{\partial s}{\partial x}=1,\hspace{0,3cm}p\frac{\partial t}{\partial y}=1,\hspace{0,3cm}q\frac{\partial t}{\partial x}=1,\hspace{0,3cm}p\frac{\partial s}{\partial y}=1, \hspace{0,3cm}p\frac{\partial q}{\partial y}\frac{\partial f}{\partial x}=q\frac{\partial p}{\partial x}\frac{\partial f}{\partial y}.
\end{equation}
To solve this set of conditions, we now make some further assumptions. As a first step, we assume that $s=t$. Then three independent conditions remain from (\ref{cond}), namely
\[
q\frac{\partial s}{\partial x}=1,\hspace{0,3cm}p\frac{\partial s}{\partial y}=1,\hspace{0,3cm}p\frac{\partial q}{\partial y}\frac{\partial f}{\partial x}=q\frac{\partial p}{\partial x}\frac{\partial f}{\partial y}.
\]
If we make in addition the assumption that $p=q$, these equations reduce to
\begin{equation}\label{cond2}
q\frac{\partial s}{\partial x}=1,\hspace{0,3cm}q\frac{\partial s}{\partial y}=1,\hspace{0,3cm}q\frac{\partial q}{\partial y}\frac{\partial f}{\partial x}=q\frac{\partial q}{\partial x}\frac{\partial f}{\partial y}.
\end{equation}
Thus, $\frac{\partial s}{\partial x}=\frac{\partial s}{\partial y}$. The structure of the equations in (\ref{cond2}) shows the following: If we choose the functions $f$ in such a fashion that $\frac{\partial f}{\partial x}=\frac{\partial f}{\partial y}$ and if we furthermore choose $q$ such that $\frac{\partial q}{\partial x}=\frac{\partial q}{\partial y}$ then all equations are satisfied provided that $q\frac{\partial s}{\partial x}=1$. Consider the functions $q(x,y):=e^{x+y}$ and $s(x,y):=e^{-(x+y)}$. They are smooth and satisfy all the conditions. If we now restrict the functions $f$ to be smooth and satisfying $\frac{\partial f}{\partial x}=\frac{\partial f}{\partial y}$, then all conditions of (\ref{cond2}) and, therefore, of (\ref{cond}) are satisfied. Thus, we have found a particular representation of the commutation relations (\ref{inttwobos}). Let us summarize the above observations in the following theorem.

\begin{Theorem}\label{theoremrep} Let $D:=\{f\in C^{\infty}(\mathbf{R}^2)\,|\, \frac{\partial f}{\partial x}=\frac{\partial f}{\partial y}\}$ and let
\begin{equation}\label{rep2}
a=e^{x+y}\frac{\partial}{\partial x}, \hspace{0,3cm}a^{\dag}=e^{-(x+y)},\hspace{0,3cm}b=e^{x+y}\frac{\partial}{\partial y}, \hspace{0,3cm}b^{\dag}=e^{-(x+y)}.
\end{equation}
Then the operators $a,a^{\dag},b,b^{\dag}$ act in $D$ and satisfy the commutation relations (\ref{inttwobos}) of the coupled two-mode boson.
\end{Theorem}
\begin{proof} Formally, we have checked by the above construction that choosing these operators satisfy the commutation relations (\ref{inttwobos}). The choice of $D$ assures us that interchanging the derivatives of $f$ in the above calculation was legitimate. The final point to check is that the operators act in $D$, i.e., that $Of\in D$ for $f\in D$ and $O \in \{a,a^{\dag},b,b^{\dag}\}$. This is straightforward.
\end{proof}

\begin{Remark} With hindsight, the above result is not too astonishing. Note that the restriction $ \frac{\partial f}{\partial x}=\frac{\partial f}{\partial y}$ on the functions $f\in D$ is extremely severe, reducing the above example essentially to the single-mode case since $a^{\dag}=b^{\dag}$ and $af=bf$ for all $f\in D$ due to the restriction! Thus, it would be interesting to find other, more complicated representations of the commutation relations (\ref{inttwobos}).
\end{Remark}

Let us consider the representation given in Theorem \ref{theoremrep} further. The number operator $N_a$ is given by $N_a=a^{\dag}a$ and, therefore, represented by $\frac{\partial}{\partial x}$ since $N_af=e^{-(x+y)}e^{x+y}\frac{\partial}{\partial x}f=\frac{\partial f}{\partial x}$. The same argument shows that $N_b$ is represented by $\frac{\partial}{\partial y}$. However, since the two derivatives coincide in $D$, one may conclude for this representation that $N_a=N_b$. Stretching the physical interpretation to its limit, we see that this is some kind of ``diagonal'' representation where the numbers of the two modes always coincide. It follows, in particular, that $[N_a,N_b]=0$ so that all conclusions derived in Example \ref{excomm} hold true. Recall that we discussed in Section 2 that one has directly from the commutation relations (i.e., independent of the representation) $[N_a,N_b]=a^{\dag}b-b^{\dag}a$. Inserting (\ref{rep2}) yields $[N_a,N_b]=\frac{\partial}{\partial x}-\frac{\partial}{\partial y}$ which indeed vanishes on every $f\in D$. This is a nice consistency check.

\section{The general multi-mode boson}
In this section we generalize the preceding sections to the general case of a multi-mode boson. For the following we assume that $n$ modes are involved, i.e., there are $n$ pairs of creation and annihilation operators $a_i^{\dag},a_i$ ($1\leq i\leq n$) each satisfying (\ref{commb}) and where the cross commutation relations are given for all $1\leq i<j \leq n$ by $[a_i,a_j^{\dag}]=1, [a_i,a_j]=0$ (together with the adjoint equations $[a_j,a_i^{\dag}]=1$ and $[a_i^{\dag},a_j^{\dag}]=0$).
\begin{Definition}
The {\it coupled $n$-mode boson} is given by the annihilation and creation operators $\{a_1,a_1^{\dag},\ldots,a_n,a_n^{\dag}\}$ satisfying for all $1\leq i,j\leq n$ the commutation relations
\begin{equation}\label{commbr}
[a_i,a_j]=0,\hspace{0,3cm}[a_i^{\dag},a_j^{\dag}]=0, \hspace{0,3cm}[a_i,a_j^{\dag}]=1.
\end{equation}
\end{Definition}

A general operator function $F(a_1,a_1^{\dag},\ldots,a_n,a_n^{\dag})$ for this multi-mode boson may be interpreted as a word on the alphabet $\{a_1,a_1^{\dag},\ldots,a_n,a_n^{\dag}\}$. The {\it normal ordering} is a functional representation of multi-mode boson operator functions in which all the creation operators $a_i^{\dag}$ stand to the left of the annihilation operators $a_i$. Although it is irrelevant from a strictly logical point of view (since $a_ia_j=a_ja_i$ as well as $a_i^{\dag}a_j^{\dag}=a_j^{\dag}a_i^{\dag}$) we will assume in the following that in the normally ordered form one always has that $a_i^{\dag}$ precedes $a_j^{\dag}$ whenever $i<j$ and that the operators $a_i$ precede the operators $a_j$ whenever $i>j$. Thus, a normally ordered form consists of a sum of terms of the form
\[
(a_1^{\dag})^{i_1}(a_2^{\dag})^{i_2}\cdots(a_n^{\dag})^{i_n}a_n^{j_n}\cdots a_2^{j_2}a_1^{j_1}.
\]
We have chosen this particular form of arrangement since it is very convenient for taking adjoints (the adjoint of such a term has again this form). As above, we denote
by $\mathcal{N}[F(a_1,\ldots,a_n^{\dag})]$ the normal ordering of the function $F(a_1,\ldots,a_n^{\dag})$.

In analogy to above we now introduce \emph{$n$-coloured contractions}: An $n$-coloured contraction of the word $F(a_1,\cdots,a_n^{\dag})$ consists of
\begin{enumerate}
\item substituting $a_i=\varnothing$ and $a_j^{\dagger}=\varnothing^{\dagger}$ in the
word whenever $a_i$ precedes $a_j^{\dagger}$ for any $1\leq i,j\leq n$,
\item and deleting all the letters $\varnothing$ and
$\varnothing^{\dagger}$ in the word.
\end{enumerate}
Among all possible $n$-coloured contractions we
also include the null contraction, that is the contraction leaving the word as
it is. We define $\mathcal{C}^{(n)}(F(a_1,\cdots,a_n^{\dag}))$ to be the multiset of all $n$-coloured contractions of $F(a_1,\ldots,a_n^{\dag})$. Again, we call an $n$-coloured contraction to be of {\it degree} $r$ if precisely $r$ pairs of creation and annihilation operators are contracted.

The {\it double dot operation} arranges a word $\pi \in \mathcal{C}^{(n)}(F(a_1,\cdots,a_n^{\dag}))$ such that
\begin{enumerate}
\item the letters $a_i^{\dag}$ precede the letters $a_j^{\dagger}$ whenever $i<j$,
\item the letters $a_i$ precede the letters $a_j$ whenever $i>j$,
\item all letters $a_i^{\dag}$ precede all letters $a_j$.
\end{enumerate}

It is now possible to generalize Theorem \ref{theorem1} as follows.

\begin{Theorem}\label{theorem2} Let $F(a_1,\ldots,a_n^{\dag})$ be an operator function of the coupled $n$-mode boson (\ref{commbr}). The normally ordered form of $F(a_1,\ldots,a_n^{\dag})$ can be described with the help of $n$-coloured contractions and the double dot operation as follows:
\begin{equation}\label{normaln}
\mathcal{N}[F(a_1,\ldots,a_n^{\dag})]=F(a_1,\ldots,a_n^{\dag})=\sum_{\pi\in\mathcal{C}^{(n)}(F(a_1,\ldots,a_n^{\dag}))}:\pi:.
\end{equation}
\end{Theorem}

In analogy to the two-mode boson where we introduced the algebra $\mathcal{A}^{(2)}$ we may here introduce the corresponding unital $*$-algebra $\mathcal{A}^{(n)}$ generated by the operators $\{1,a_1,\ldots,a_n^{\dag}\}$ satisfying the above commutation relations (\ref{commbr}). Clearly, the operation of normal ordering $\mathcal{N}\equiv \mathcal{N}_n$ is a map $\mathcal{N}_n: \mathcal{A}^{(n)}\rightarrow \mathcal{A}^{(n)}$. As in the two-mode case we have here a $*$-algebra-homomorphism $\mathcal{SUN}:\mathcal{A}^{(n)}\rightarrow \mathcal{A}^{(1)}$ mapping $a_i\mapsto c$ and $a_i^{\dag}\mapsto c^{\dag}$ for $1\leq i \leq n$; we also call it ``putting the sunglasses on''. We now have the following generalization of Theorem \ref{sunglass2}:

\begin{Theorem}\label{gensun}
Let $n$ be an arbitrary integer. Consider the algebras $\mathcal{A}^{(n)}$ defined above and let us denote the operation of normal ordering in these algebras by $\mathcal{N}_n:\mathcal{A}^{(n)}\rightarrow \mathcal{A}^{(n)}$. Then the operations of normal ordering and ``putting the sunglasses on'' commute, i.e.,
\begin{equation}
\mathcal{SUN}\circ \mathcal{N}_n =\mathcal{N}_1\circ\mathcal{SUN}.
\end{equation}
\end{Theorem}

We may also introduce in close analogy to Section \ref{sectwostirl} the general {\it $n$-coloured Stirling numbers} of both types as follows.
\begin{Definition}
Let $(n_1,\ldots,n_n)\in \mathbf{N}^n, m\in \mathbf{N}$ and assume that the operators $\{a_i,a_i^{\dag}\}$ for $1\leq i \leq n$ satisfy the commutation relations (\ref{commbr}). The $n$-{\it coloured Stirling numbers of first type} $S_{n_1,\ldots,n_n}(i_1,\ldots, i_{n},j_n,\ldots,j_1)$ are defined by
\begin{equation}
\mathcal{N}[N_{a_1}^{n_1}\cdots N_{a_n}^{n_n}]=\sum_{i_1,\ldots,i_{n} \atop j_1,\ldots,j_{n} }S_{n_1,\ldots,n_n}(i_1,\ldots, i_{n},j_n,\ldots,j_1)(a_1^{\dag})^{i_1}\cdots(a_n^{\dag})^{i_n}a_n^{j_n}\cdots a_1^{j_1}
\end{equation}
and the {\it $n$-coloured Stirling numbers of second type} $S_{n}(i_1,\ldots, i_{n},j_n,\ldots,j_1)$ are defined by
\begin{equation}
\mathcal{N}[(N_{a_1}\cdots N_{a_n})^m]=\sum_{i_1,\ldots,i_{n} \atop j_1,\ldots,j_{n}}S_{m}(i_1,\ldots, i_{n},j_n,\ldots,j_1)(a_1^{\dag})^{i_1}\cdots(a_n^{\dag})^{i_n}a_n^{j_n}\cdots a_1^{j_1}.
\end{equation}
\end{Definition}
These $n$-coloured Stirling numbers can then be discussed like the ones of the two-coloured case (explicit values, recursion relations, etc.). Note that we could also define less symmetric Stirling numbers which result by ``clustering'' some of the modes. From this point of view the $n$-coloured Stirling numbers of first type do not cluster the modes, i.e., each mode has its own exponent whereas the $n$-coloured Stirling numbers of second type are the opposite case since there is one exponent for all modes. In between are the aforementioned less symmetric versions. For example, one could consider $\mathcal{N}[(N_{a_1}\cdots N_{a_k})^{l_1}(N_{a_{k+1}}\cdots N_{a_n})^{l_2}]$ and define associated Stirling numbers as above. In analogy to the two-mode case one can also introduce the associated $n$-{\it coloured Bell numbers of first resp. second type} by
\begin{eqnarray*}
B^{(n)}(n_1,\ldots,n_n)&:=&\sum_{i_1,\ldots,i_{n} \atop j_1,\ldots,j_{n} }S_{n_1,\ldots,n_n}(i_1,\ldots, i_{n},j_n,\ldots,j_1),\\ B^{(n)}(m)&:=&\sum_{i_1,\ldots,i_{n} \atop j_1,\ldots,j_{n}}S_{m}(i_1,\ldots, i_{n},j_n,\ldots,j_1).
\end{eqnarray*}
Using Theorem \ref{gensun}, one may show the analogue of Theorem \ref{Bellcomp}:
\[
B^{(n)}(n_1,\ldots,n_n)=B(n_1+\cdots+n_n), \hspace{0,3cm} B^{(n)}(m)=B(nm).
\]

Let us now introduce the general $n$-{\it coloured linear representations} to depict $n$-coloured contractions, thereby generalizing the construction of Section \ref{tclr}.  We assume that $n$ colours $c_i$ associated to mode $a_i$ for $1\leq i \leq n$ have been chosen. Let us consider a word $\pi$ on the alphabet $\{a_1,\ldots,a_n^{\dag}\}$ of length $m$, i.e., $\pi=\pi_1\cdots \pi_m$ with $\pi_j\in \{a_1,\ldots,a_n^{\dag}\}$. We
draw $m$ vertices, say $1,2,\ldots,m$, on a horizontal line, such that the
point $j$ corresponds to the letter $\pi_{j}$. We represent
\begin{enumerate}
\item each letter $a_i$ by an empty circle of colour $c_i$,
\item each letter $a_i^{\dagger}$ by a filled circle of colour $c_i$.
\end{enumerate}
A filled circle $j$ can be connected by an undirected edge $(i,j)$ to an empty circle $i$ if $i<j$ - independent of the colours of the circles. As above, the edges are drawn in the plane above the points. This is the $n$-coloured linear representation of an $n$-coloured contraction.

A generalization of the representation discussed in Theorem \ref{theoremrep} to the case of $n$ modes is easy to find:
\begin{Theorem} Let $D_n:=\{f\in C^{\infty}(\mathbf{R}^n)\,|\,\frac{\partial f}{\partial x_i}=\frac{\partial f}{\partial x_j}\mbox{ for } 1\leq i,j\leq n\}$ and define the operators $a_i,a_i^{\dag}$ for $1\leq i\leq n$ by
\[
a_i=e^{x_1+\cdots+x_n}\frac{\partial}{\partial x_i}, \hspace{0,3cm}a_i^{\dag}=e^{-(x_1+\cdots+x_n)}.
\]
Then the operators $a_i,a_i^{\dag}$ for $1\leq i\leq n$ act in $D_n$ and satisfy the commutation relations (\ref{commbr}) of the coupled $n$-mode boson.
\end{Theorem}

\section{Conclusions}
In this paper we began discussing the combinatorial problem of normal ordering words in the annihilation and creation operators of the coupled two-mode boson. It was shown that this problem cannot be reduced in a straightforward way to the well-known case of a single-mode boson. On the more conceptual side we have introduced two-coloured contractions and have shown how the normally ordered form of an arbitrary word in the creation and annihilation operators can be expressed as a sum over the two-coloured contractions, generalizing the corresponding result for the single-mode case. We also described the two-coloured linear representations as a nice depiction of two-coloured contractions. In the single-mode case one has the beautiful observation of Katriel that normal ordering powers of the number operator involves the Stirling numbers of second kind. In view of this we were led to define in the two-mode case the two-coloured Stirling numbers as coefficients appearing when normal ordering powers of the number operators of the two-mode case. These two-coloured Stirling numbers are a natural generalization of the conventional Stirling numbers and share several properties with them. We derived several of these properties, in particular recursion relations and some explicit values. Clearly, a more detailed study of these numbers and their combinatorial interpretation is desirable. The associated two-coloured Bell numbers were introduced and it was shown that they can be expressed through the conventional Bell numbers. All of the above aspects of the two-mode boson can be generalized to the case of $n$ modes. We sketched most of these points in the last section where we introduced $n$-coloured contractions, $n$-coloured linear representations and $n$-coloured Stirling numbers and showed how the results mentioned above for the two-mode case can be generalized to the $n$-mode case. Apart from the points already mentioned it would clearly be interesting to find nontrivial representations for the commutation relations of the coupled $n$-mode boson, in particular Fock-space like representations.

\subsection*{Acknowledgements}
The authors would like to thank Simone Severini for useful discussions.

\end{document}